\documentclass{ws-procs975x65}

\begin{document}

\title{\uppercase{Gravitational waveforms for unequal mass black hole binaries detectable by KAGRA} 
\footnote{Work supported by the European Union / European Social Fund grants T\'{A}MOP-4.2.2.A-11/1/KONV-2012-0060, T\'{A}MOP-4.2.2/B-10/1-2010-0012, 
T\'{A}MOP-4.2.2.C-11/1/KONV-2012-0010 and OTKA grant no. 100216.}}

\author{\uppercase{M\'{a}rton T\'{a}pai}$^{\dag}$, \uppercase{Zolt\'{a}n Keresztes} and \uppercase{L\'{a}szl\'{o} \'{A}. Gergely}}

\address{Departments of Theoretical and Experimental Physics, University of Szeged,\\
Szeged, 6720 D\'{o}m t\'{e}r 9., Hungary\\
$^{\dag}$ E-mail: tapai@titan.physx.u-szeged.hu}

\begin{abstract}
Gravitational waveforms generated by unequal mass black hole binaries are
expected to be common sources for future gravitational wave detectors. We
derived the waveforms emitted by such systems during the last part of the 
inspiral, when the larger spin dominates over the orbital angular 
momentum and the smaller spin is negligible. These Spin-Dominated 
Waveforms (SDW) arise as a double expansion in the post-Newtonian parameter
and another parameter proportional to the ratio of the orbital angular momentum and the dominant spin. The time spent by the gravitational wave as an SDW in the sensitivity range of the KAGRA detector is presented for the first time.
\end{abstract}


\bodymatter

\section{Introduction}
Ground based gravitational detectors, the Advanced LIGO (aLIGO), the Einstein Telescope and the Large-scale cryogenic gravitational wave detector (KAGRA) also the space missions LISA and LAGRANGE will detect gravitational waves (GW) from black hole binaries. For astrophysical black hole binaries there is no preferred mass ratio $\nu=m_{2}/m_{1}$, and for supermassive black hole binaries typically $\nu \in \left[ 0.3 , 0.03 \right]$\cite{spinflip1}.
In this regime, at the end of the inspiral, the larger spin dominates over the orbital angular momentum, represented by the smallness of the parameter $\xi =\varepsilon ^{-1/2}\nu $. Here $\varepsilon =Gm/c^{2}r\approx v^{2}/c^{2}$ is the post-Newtonian (PN) parameter (with the orbital
separation $r$, the orbital velocity $\ v$ of the reduced mass particle $\mu
=m_{1}m_{2}/m$, the gravitational constant $G$, the speed of light $c$). Hence the corresponding spin-dominated waveforms (SDW) are simpler. The time spent by the GW as SDW in the sensitivity ranges of aLIGO, Einstein Telescope, LISA and LAGRANGE were presented in Ref.~\refcite{SDW}, while for eLISA in Ref.~\refcite{prague}. In this paper the respective time for the KAGRA detector will be also given.

\section{Spin-dominated waveforms}
For spinning black hole binaries, the ratio of the smaller and larger spin magnitude is of order $S_{2}/S_{1} = \nu^2\chi_{2}/\chi_{1}$\cite{spinflip1} (where $\chi_{i}$ represent the dimensionless larger spin, taken here close to the maximally allowed limit 1). Thus to first order in $\nu$, the smaller spin can be neglected. The ratio of the orbital angular momentum $L_{N}$ and $S_{1}$ is $L_{N}/S_{1} \approx \varepsilon ^{-1/2}\nu \chi^{-1}_{1}$. \cite{spinflip1}

Generic waveforms to 1.5 PN order\cite{kidder,ABFO}, and 2 PN order\cite{BFH} accuracy were previously known. We employed a second taylor expansion in the parameter $\xi$. The obtained SDW has the following structure\cite{SDW}: 
\begin{eqnarray}
h_{_{\times }^{+}} &=&\frac{2G^{2}m^{2}\varepsilon ^{1/2}\xi }{c^{4}Dr}\left[
h_{_{\times }^{+}}^{0}+\beta _{1}h_{_{\times }^{+}}^{0\beta }+\varepsilon
^{1/2}\left( h_{_{\times }^{+}}^{0.5}+\beta _{1}h_{_{\times }^{+}}^{0.5\beta
}-2\xi h_{_{\times }^{+}}^{0}\right) +\varepsilon \left( h_{_{\times
}^{+}}^{1}-4\xi h_{_{\times }^{+}}^{0.5}\right. \right.   \notag \\
&&\left. +\beta _{1}h_{_{\times }^{+}}^{1\beta }+h_{_{\times
}^{+}}^{1SO}+\beta _{1}h_{_{\times }^{+}}^{1\beta SO}\right) \left.
+\varepsilon ^{3/2}\left( h_{_{\times }^{+}}^{1.5}+h_{_{\times
}^{+}}^{1.5SO}+h_{_{\times }^{+}}^{1.5tail}\right) \right] ~,
\end{eqnarray}
D being the luminosity distance of the source. The $\xi$ and $\varepsilon$ orders of the contributions to $h_{_{\times }^{+}}$ are given in Table \ref{table1}.
\begin{table}
\tbl{SDW contributions of different $\xi$ and $\varepsilon$ orders. The SO terms contain the dominant spin.}
{\begin{tabular}{@{}ccccc@{}}
\toprule
& $\varepsilon ^{0}$ & $\varepsilon ^{1/2}$ & $\varepsilon ^{1}$ & $%
\varepsilon ^{3/2}$ \\
\colrule
$\xi ^{0}$ & $h_{_{\times }^{+}}^{0}$ & $h_{_{\times }^{+}}^{0.5}$ & $%
h_{_{\times }^{+}}^{1},h_{_{\times }^{+}}^{1SO}$ & $h_{_{\times
}^{+}}^{1.5},h_{_{\times }^{+}}^{1.5SO},h_{_{\times }^{+}}^{1.5tail}$ \\ 
$\xi ^{1}$ & $h_{_{\times }^{+}}^{0\beta }$ & $h_{_{\times }^{+}}^{0.5\beta
} $ & $h_{_{\times }^{+}}^{1\beta },h_{_{\times }^{+}}^{1\beta SO}$ &  \\ 
\botrule
\end{tabular}
}
\label{table1}
\end{table}

The phase of the gravitational waveform in the double expansion becomes:\cite{SDW}
\begin{eqnarray}
\phi _{c}-\phi &=&\frac{\varepsilon ^{-3}}{32\xi }\left\{ 1+2\varepsilon
^{1/2}\xi +\frac{1195}{1008}\varepsilon  +\left( \allowbreak -10\pi +\frac{3925}{504}\xi +\frac{175}{8}\chi_{1}\cos \kappa _{1}\right) \varepsilon ^{3/2} \right.  \notag \\
&& \left. +\left[ -\frac{21\,440\,675}{1016\,064}  +\chi _{1}^{2}\left( \frac{375}{16}-\allowbreak \frac{3425}{96} \sin ^{2}\kappa _{1}\right) \right] \varepsilon ^{2}\allowbreak \right\} ~.
\end{eqnarray}
Here $\phi _{c}$ is the phase at the coalescence, and $\kappa_{1}$ is the angle span by $S_{1}$ and $L_{N}$.

\section{Limits of Validity}
\begin{figure}[t]
\begin{center}
\psfig{file=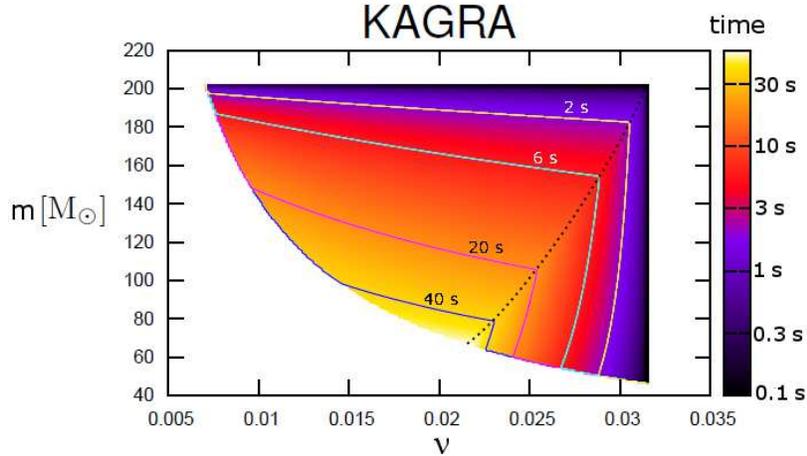,width=4.5in}
\end{center}
\caption{The time during which SDWs are detectable by KAGRA is represented as function of the total mass $m$ and mass ratio $\nu$. The color code is logarithmic.}
\label{fig1}
\end{figure}

We impose $\xi \leq 0.1$, equivalent to a lower limit $\varepsilon _{1}=Gm/c^{2}r_{1}=100\nu ^{2}$ for the PN parameter. The end of the inspiral gives an upper limit for the PN parameter, chosen here as $\varepsilon _{2}=0.1$\cite{LEVIN}. This leads to an upper limit for the mass ratio $\nu_{\max }=0.0316\approx 1:32$.
The time during which the binary evolves from $\varepsilon _{1}$ to $\varepsilon _{2}$ is\cite{SDW}
\begin{equation}
\Delta t=\frac{5Gm}{2^{8}c^{3}}\frac{(1+\nu )^{2}}{\nu }\left( \varepsilon
_{1}^{-4}-\varepsilon _{2}^{-4}\right) ~.
\end{equation}
For a given SDW, $\varepsilon_{1}$ can be lower than the value of the PN parameter at the lower sensitivity bound $\varepsilon_{f_{\min }}$, hence the 
time spent in the best sensitivity range of the detector by an SDW is calculated from $\max \left( \varepsilon_{1} , \varepsilon_{f_{\min }}\right) $ 
to $\varepsilon _{2}$.
We represented on Fig \ref{fig1} this time interval for the KAGRA detector, as function of $m$ and $\nu$. The lower sensitivity bound of the KAGRA 
detector is $f_{min}=10 Hz$\cite{KAGRA}, same as for aLIGO \cite{aLIGO}. The corresponding figure for aLIGO is Fig 1 of Ref.~\refcite{SDW}. Although the 
figures are similar, the detectors exhibit different shapes of the spectral noise density as function of frequency, leading to different signal-to-noise ratios for a waveform.

We derived the upper limit of the total mass $m = 202~$M$_{\odot }$ from the lower frequency bound of the KAGRA. By assuming the smaller compact object has at least the mass of a neutron star ($1.4~$M$_{\odot }$) we found a total mass dependent lower limit $\nu_{\min}$, represented by the lower cutoff on Fig \ref{fig1}.

\section{Concluding Remarks}
For mass ratios smaller than $\nu_{\max}=1:32$ the larger spin $S_{1}$ dominates over the orbital angular momentum $L_{N}$ at the end of the inspiral. This is expressed by a second small parameter $\xi$ (beside the PN parameter $\varepsilon$). Expanding the PN waveforms in terms of $\xi$ (to first order, this also leads to the neglection of the secondary spin) leads to the spin-dominated waveforms.
These waveforms are considerably shorter than the generic waveforms. The corresponding smaller parameter space will turn advantageous in gravitational wave detection.

\bibliographystyle{ws-procs975x65}
\bibliography{mtapai_mg13}

\end{document}